\documentclass[narrowdisplay]{elsarticle}
\usepackage{hyperref}
\usepackage{amssymb}
\journal{Nuclear Instruments and Methods in Physics Research~~~}

\begin{document}

\begin{frontmatter}

\title{A New Technique for Large-Area  Detection of High Energy Particles using Ultra-Fast Magnetic Sensing}%\tnotetext[mytitlenote]{Fully documented templates are available in the elsarticle package on \href{http://www.ctan.org/tex-archive/macros/latex/contrib/elsarticle}{CTAN}.}

%% Group authors per affiliation:
\author{David Saltzberg and Peihao Sun}
\address{Dept. of Physics and Astronomy, University of California, Los Angeles\\ Los Angeles, California 90095-1547 USA}

%%% or include affiliations in footnotes:
%\author[mymainaddress,mysecondaryaddress]{Elsevier Inc}
%\ead[url]{www.elsevier.com}

%\author[mysecondaryaddress]{Global Customer Service\corref{mycorrespondingauthor}}
%\cortext[mycorrespondingauthor]{Corresponding author}
%\ead{saltzberg@physics.ucla.edu}

%\address[mymainaddress]{1600 John F Kennedy Boulevard, Philadelphia}
%\address[mysecondaryaddress]{360 Park Avenue South, New York}

\begin{abstract}
Cascades from high-energy particles produce a brief current and associated magnetic fields. Even sub-nanosecond duration magnetic fields can be detected with a relatively low bandwidth system by latching image currents on a capacitor.  At accelerators, this technique is employed routinely by beam-current monitors, which work for pulses even as fast as femtoseconds.   We discuss scaling up these instruments in size, to 100 meters and beyond, to serve as a new kind of ground- and space-based high-energy particle detector which can instrument large areas relatively inexpensively.  This new technique may be used to detect and/or veto ultra-high energy cosmic-ray showers above 100~PeV. It may also be applied to searches for hypothetical highly charged particles.  In addition, these detectors may serve to search for extremely short magnetic field pulses of any origin, faster than other detectors by orders of magnitude. 
\end{abstract}

\begin{keyword}
\texttt{instrumentation:detectors} \sep \texttt{magnetic fields}  \sep \texttt{ultra high energy cosmic rays} 
%\MSC[2010] 00-01\sep  99-00
\end{keyword}

\end{frontmatter}

%\linenumbers

\section{Introduction}

Maxwell's equations relate any current that crosses a two-dimensional surface to a one-dimensional magnetic line integral around its perimeter.  This reduction in dimension suggests an efficient means for particle detection over large areas requiring instrumentation over only one dimensional contours.   This relationship is already exploited regularly over small areas by accelerator physicists.   In this paper we explore taking the known techniques from the existing scale of 0.1~m and expanding them to a 100~m scale, which would be of interest to particle astrophysicists.

The common feature of the accelerator physics techniques we borrow from is a toroidal geometry.  Once a current passes through the plane of a detector, closed loops of magnetic field are created, including inside a toroid constructed of conductive material.   Currents running on the toroid flow to cancel changing magnetic fields within its volume.  It is a remarkable feature that when certain basic conditions are met, the net current that flows on the surface of the toroid equals the initiating current, independent of the radius of the toroid.

In this paper we derive the conditions for detectability of fast pulses of relativistic charges by large toroids.   We calculate thresholds and demonstrate detectability of cosmic ray air showers partly for their own interest but also as a calibration and proof-of-principle.   We will note that there is an experimentally unexplored regime of area versus sampling fraction which this new technique can address.

\section{Integrated Detection}

\begin{figure}[t]
\begin{center}
\includegraphics[width=4in]{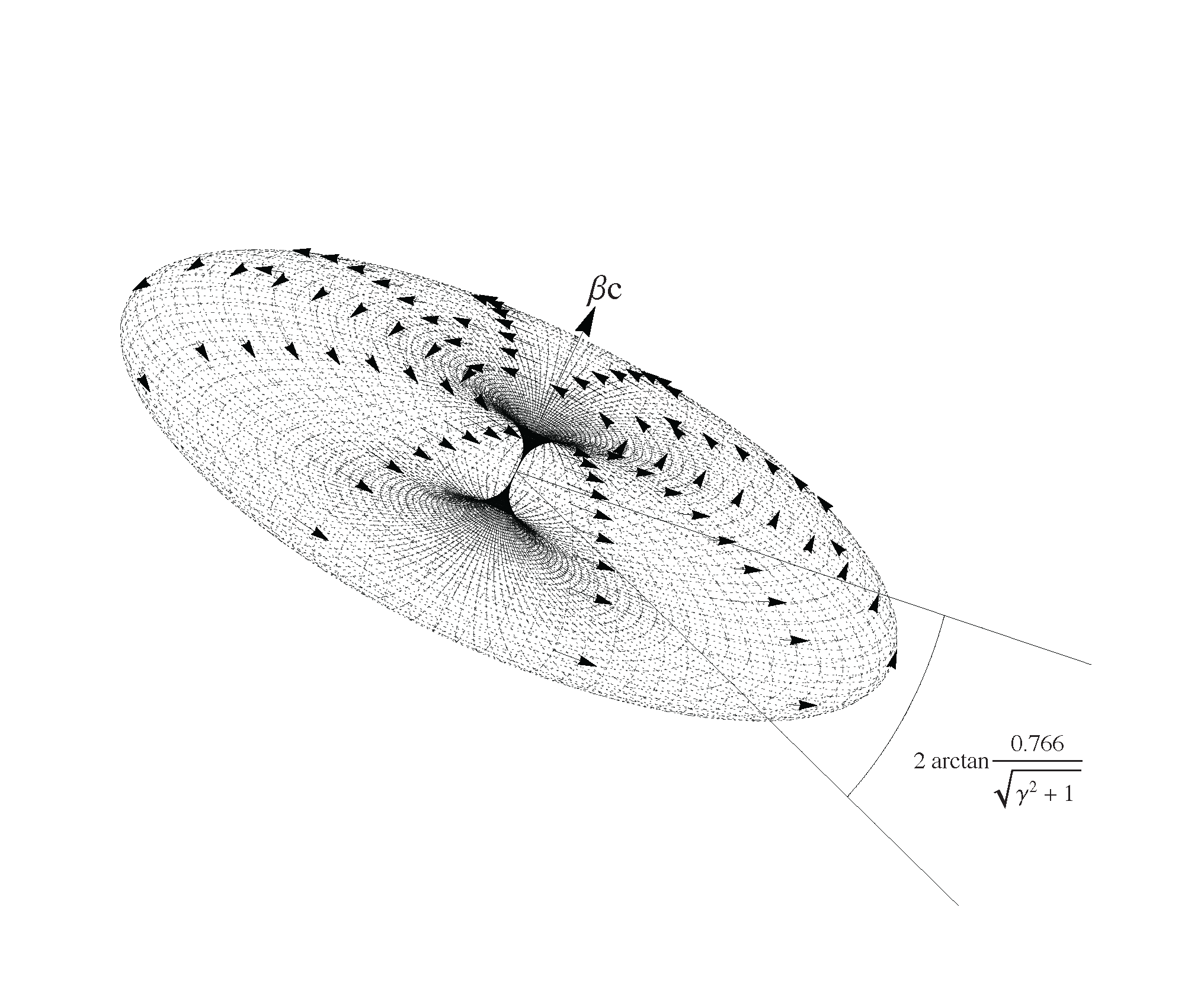}
\end{center}
\caption{Surface of constant B-field intensity for a positive particle with speed $\beta c$. Arrows indicate the B field direction.  For clarity, the surface is drawn for a particle with speed corresponding to only $\gamma=\sqrt{10}$.  For the applications considered here, $\gamma$ ranges from 100  for shower maximum of cosmic rays to 20,000 for typical beam energies.  In both cases, the surface will be much flatter than shown. \label{B-drawing}}
\end{figure}

As shown in Figure~\ref{B-drawing}, the magnetic field of a charged particle is relativistically contracted into a small solid angle oriented perpendicularly to the direction of motion.  The opening angle is inversely proportional to the Lorentz factor, {\it i.e.}, $1/\gamma$, so that at a distance $r$ from the axis of motion, the magnetic field has a duration $r/(c\gamma$), where $c$ is the speed of light.   Particles at shower maximum in a cosmic-ray air shower have energies of order 50-100~MeV. Hence 
$\gamma$ is of order 100--200, making the duration of the magnetic pulse ranging from fractions of a nanosecond to several nanoseconds at distances up to 100~meters.   
When this technique is applied to accelerators with 10~GeV electrons, measured at distances of about 0.1~m, the duration is even shorter, of order 20~femtoseconds, if not limited by the longitudinal beam bunch duration itself, typically picoseconds.  For $10^6$ electrons of energy 75~MeV  at a distance of 100~m, the peak magnetic field is $\sim10^{-13}$~Tesla.
%eg http://farside.ph.utexas.edu/teaching/em/lectures/node125.html formula 1539

\begin{figure}
\begin{center}
\includegraphics[width=5in]{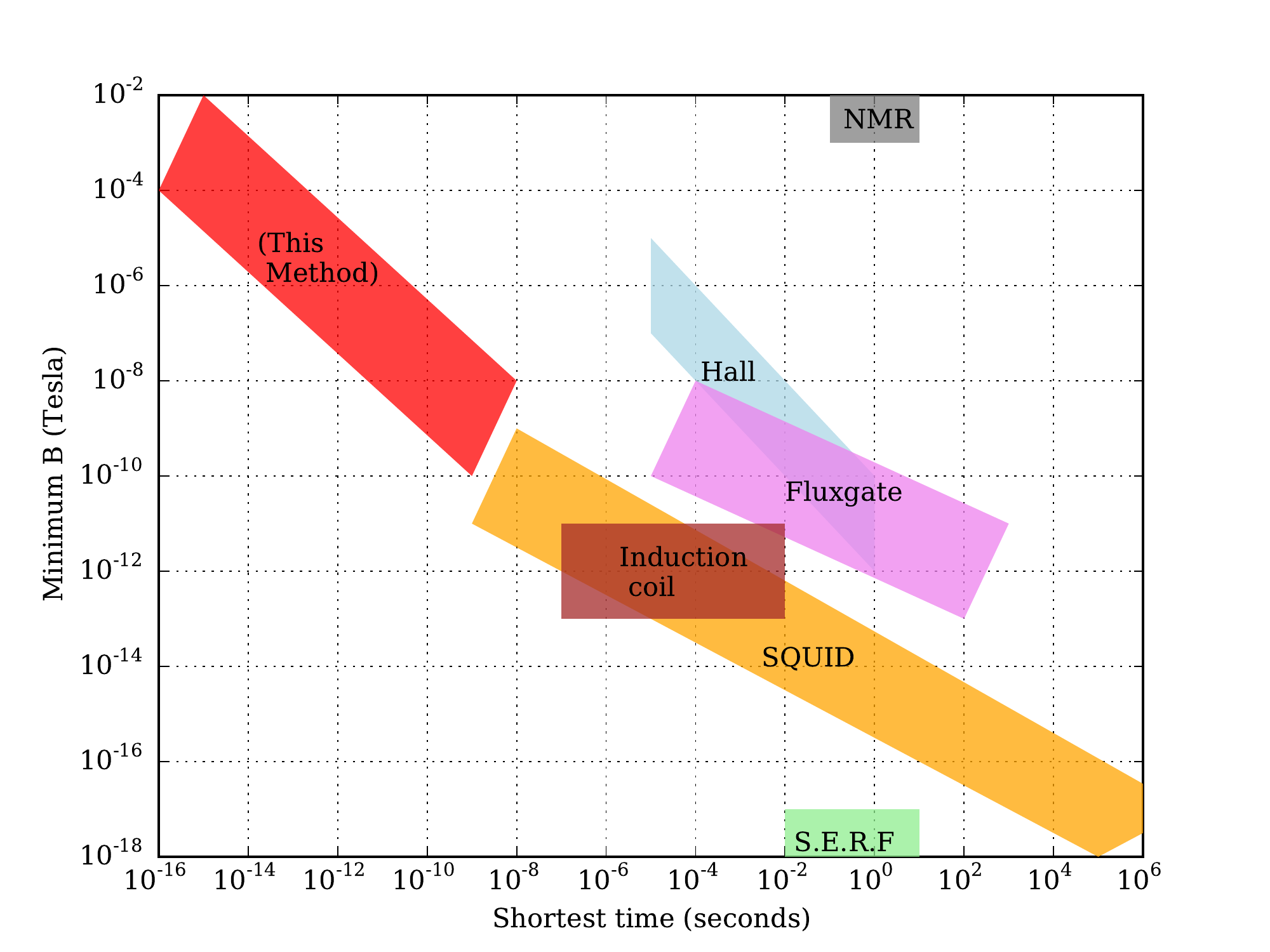}
\end{center}
\caption{Sensitivity to magnetic fields for various technologies vs integration or sampling time.  The red box is the technology considered in this paper, latched induction, which is the basis of an ICT.\label{b-vs-t}}
\end{figure}

As shown in Figure~\ref{b-vs-t}, the sensitivity of real-world magnetic field measurements depends on the duration over which they can be measured.  The well-known highest sensitivity techniques, such as SQUIDs, are not applicable to these short durations.  Even the fastest shown techniques, induction coils, would require electronics with at least GHz bandwidth to detect these fast pulses.  At accelerators the bandwidth difficulty is even worse, a direct measurement requiring THz or even PHz bandwidth.  However, the rate of pulses expected at accelerators or from cosmic rays ranges from milliHertz to microHertz or below --- suggesting that such high or impossibly high bandwidth electronics are not  necessary.

\begin{figure}[p]
\begin{center}
\includegraphics[width=3.5in]{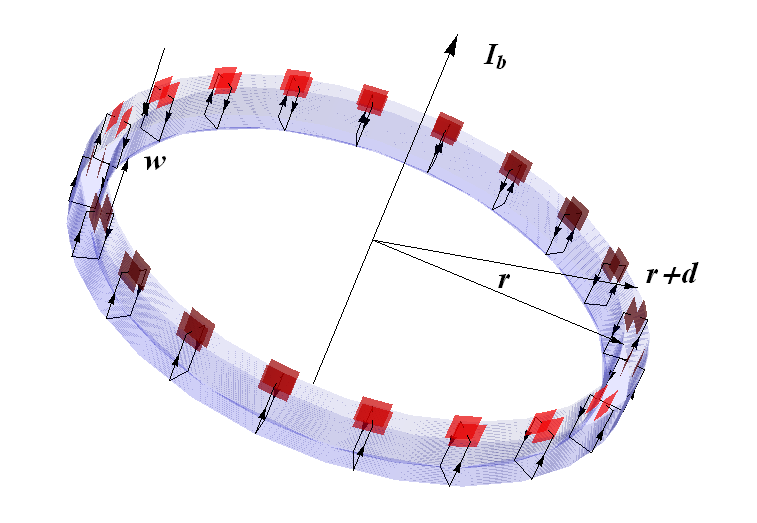}
\end{center}
\caption{Conductive toroid for detection of relativistic charges, represented by a beam current, $I_b$.  The inner radius is $r$ and the outer radius is $r+d$. The width along the beam direction is $w$.  The induced currents  flow antiparallel to the beam on the inner surface.  The red plates represent capacitors in parallel.  \label{toroid}}
\end{figure}

\begin{figure}[p]
\begin{center}
\includegraphics[width=3.8in]{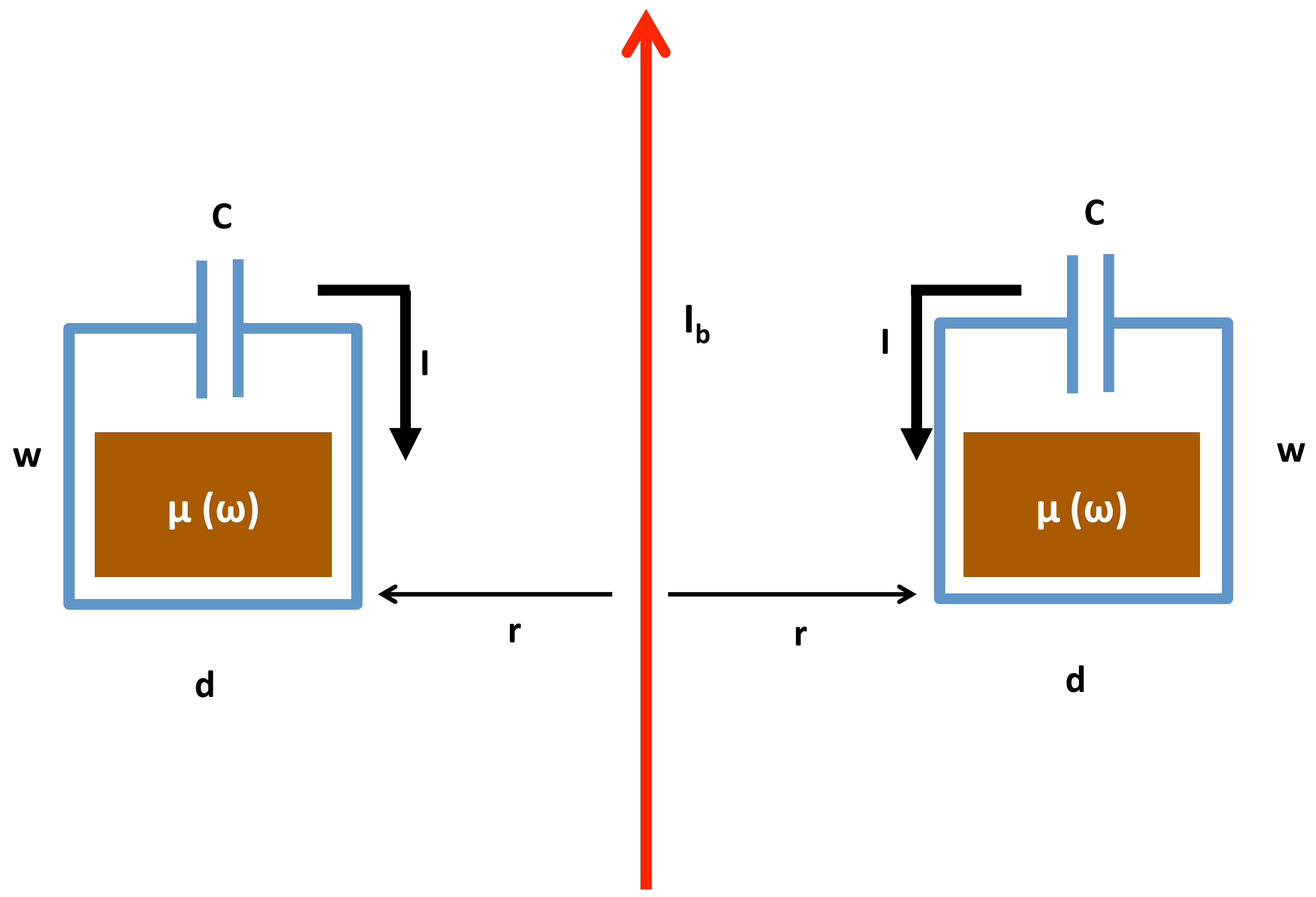}\\
\end{center}
\caption{Side, schematic, view of the toroid shown in Figure~\ref{toroid}.  Also shown is the magnetic material inside with frequency-dependent permeability, $\mu(\omega)$.  \label{toroid-side-view}}
\end{figure}

This problem has long been solved at accelerators using the so-called ``Integrating Charge Transformer'' (ICT)~\cite{unser89, bergozwebsite} and a variation of the well-known wall-current monitor known as a ``Wall Charge Monitor'' (WQM).~\cite{lebedev08}   First, the beam bunch charge traverses the plane of a toroid as shown in Figures~\ref{toroid} and~\ref{toroid-side-view}.   In vacuum, these relativistic charges produce an azimuthal magnetic field as in Figure~\ref{B-drawing}.  Then, because the toroid is made of conductive material,  currents will flow on the toroid, anti-parallel to beam current  on the inner wall (and parallel on the outer wall).  The magnitude of these currents is given by Lenz's law so that they cancel the magnetic field that would have been produced inside the toroid.  Under the conditions derived in Subsection~\ref{subsect:chargingphase}, the net charge that flows equals the charge, $Q_\mathrm{b}$, that just passed through the plane of the device, where $b$ stands for beam or any bunch of charge.  This is the charge which will be stored on the capacitor as well.  However, the toroid itself is filled with material with high magnetic permeability so that the capacitor cannot discharge as quickly as it was charged.  A typical time constant for the discharge in particle-beam applications is about 50~nanoseconds with the exact shape and duration determined by the inductance, capacitance and resistance as described below.  An ICT measures the charge on the capacitor from an integral of its discharge measured through a transformer.  The WQM measures the charge stored on the capacitor directly from its voltage.
The best such devices are able to measure beam charges as small as approximately 100~fC (625,000 electrons).~\cite{bergozwebsite}

We show below that the induced charge on the capacitors is independent of the size of the toroid.  While for larger radii, the magnetic field and hence corresponding inductance is lower, the same net charge in response to a passing charge must appear on the capacitor.   In this paper we determine the modifications required to scale up the detector size and analyze its sensitivity.

\section{Analysis and Simulation}

\begin{figure}
\begin{center}
\includegraphics[width=4in]{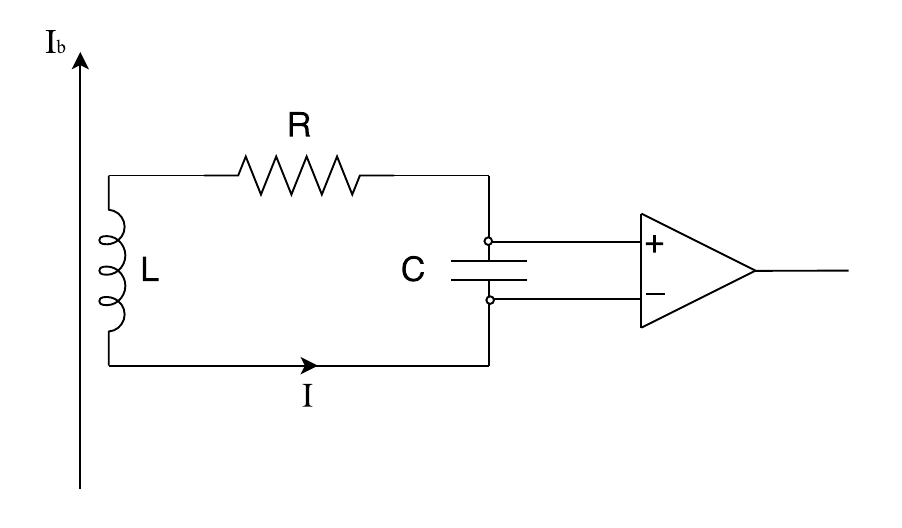}
\end{center}
\caption{Equivalent circuit for the apparatus.  The beam serves as the primary winding of a transformer whose secondary winding is the inductor, $L$.  The amplifier measures the voltage (charge) on the capacitor.  \label{equivalent-circuit}}
\end{figure}

The equivalent circuit for the device is shown in Figure~\ref{equivalent-circuit}.
We analyze the circuit in two steps, the charging and discharging phases.    We need to analyze them separately because $R$ and $L$ will be different during the two phases since the time scales are different.  Also during the first phase we are driving the circuit with a net emf around the loop and there is no such emf during the discharge.

For each phase, the toroid may be analyzed as a series $RLC$ circuit with an induced current $I(t)$. The storage capacitor has capacitance $C$ and the material of the toroidal shell presents a resistance 
$R$.  Losses in the dielectric and magnetic materials that lower the quality factor, QF, of the circuit are describable by increasing $R$.
The toroid has a mutual inductance with the beam, $M$, and self-inductance,  $L$.
The current in the circuit versus time $t$ is given by
\begin{equation}
RI(t) +L\frac{dI(t)}{dt} +\frac{1}{C}\int_{-\infty}^{t}I(t')dt' = M\frac{dI_b(t)}{dt}
\label{maindiffeqn}
\end{equation}
where $I_b(t)$ describes the pulse of beam current and is zero in the discharging phase.   
%We are only considering here the cases where the duration of the beam pulse is much less than any characteristic times of the toroid. 

During the charging phase, the resistance $R$ might have been significant since for fast pulses of current in the toroidal shell, the current flows through a small cross sectional area given by a short skin depth, $\delta$. This can be calculated using the dimensions of the toroid shown in Figures~\ref{toroid} and~\ref{toroid-side-view}, where $r$ is its inner radius, $r+d$ is its outer radius and $w$ is its width, or extent along the beam axis.    The resistance is approximately $R=\rho \ell/A$ where the path length, $\ell$, around the toroid is $2(w+d)$ and the area, $A$, is approximately $2\pi r \delta$.      We will see later that this $R$ is small compared to other magnetic and dielectric losses.

Note that during the charging phase, $L$ is smaller than it will be during the discharging phase since the relative magnetic permeability at high frequencies is close to unity, {\it i.e.}, $\mu_r=1$ rather than $\mu_r \sim 1000$.

\subsection{Charging Phase}
\label{subsect:chargingphase}
In the frequency domain the solution to Equation~\ref{maindiffeqn} is
\begin{equation}  \label{eqn:I I_b}
I(\omega)= \frac{i\omega M}{Z_L+R+Z_C} I_b(\omega),
\end{equation}
where $\omega$ is angular frequency and $Z_L$ and $Z_C$ are the impedances of the inductor and capacitor, respectively.

Using $Z_L=i\omega L$ and $Z_C=-i/(\omega C)$ we have:
\begin{equation}
I(\omega)= \frac{i\omega M}{i\omega L+R-\frac{i}{\omega C}} I_b(\omega).
\end{equation}
As shown in~\ref{inductances}, for a closed loop, $M$ equals $L$ and the toroid's self inductance is $L=\frac{\mu_r \mu_0 w}{2\pi} \ln (\frac{r+d}{r})$ which we use to calculate $Z_L$.   We can maximize $I$ for all frequencies when $|Z_L|\gg R$ and $|Z_L|\gg |Z_C|$.   As expected, we see $I=I_b$.  

We examine these conditions on the impedances for several different-sized toroids.   
First, we consider a hypothetical ICT being used for a 10~picosecond beam pulse.   Let the toroid have inner and outer radii of  $r=0.05$~m and $r+d=0.06$~m and a width $w=0.025$~m.  It is fabricated from copper which has a skin depth at 100~GHz of 200~nm.  A typical capacitance with which such devices are fabricated is 4000~pF.  In this case, $|Z_L|=12000\ \Omega$, to be compared to the resistance of $0.2\ \Omega$ and 
$|Z_C|= 0.0025\ \Omega$, so the conditions above are well met.

\begin{figure}
\begin{center}
\includegraphics[height=4in]{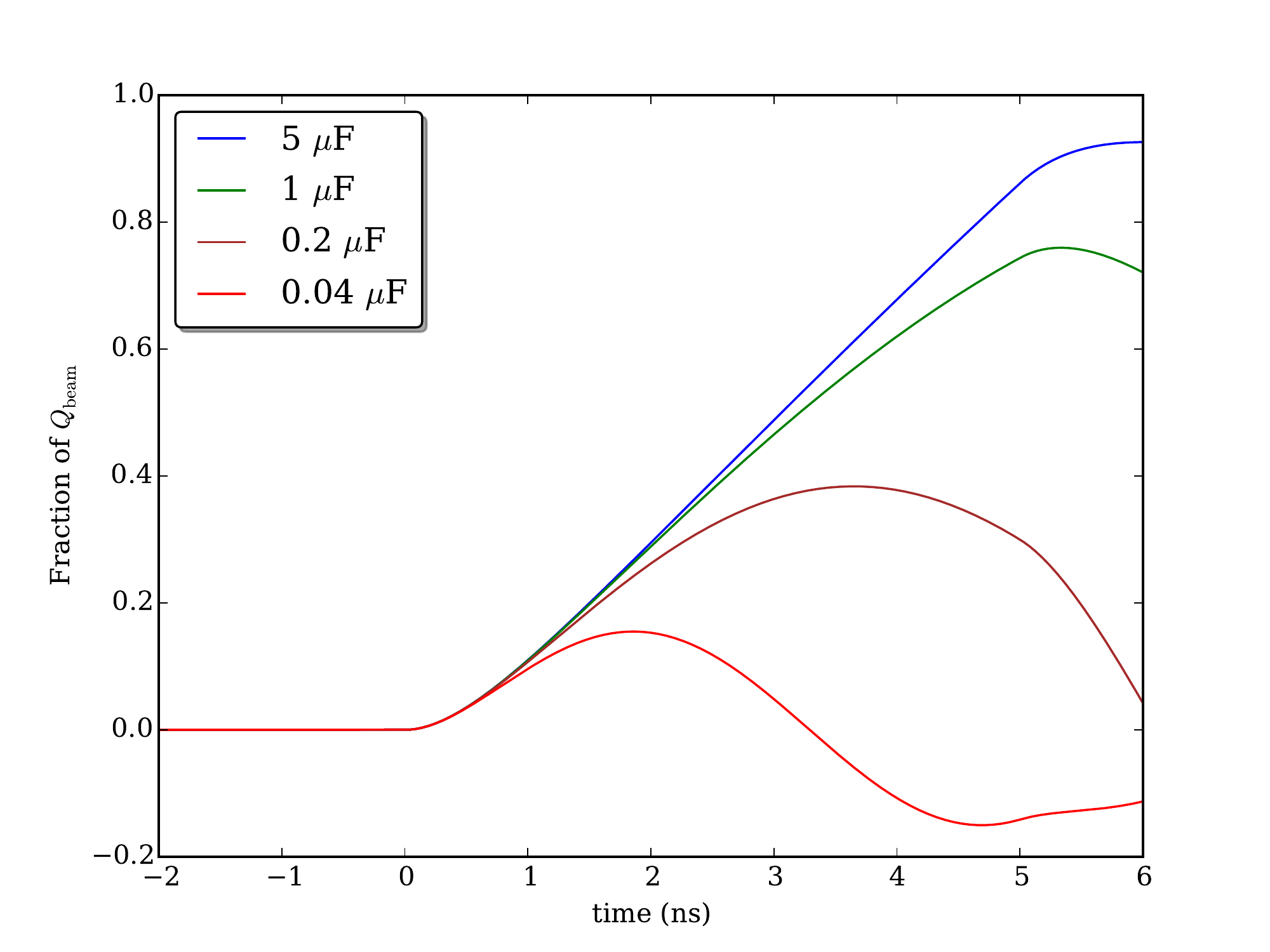}
\end{center}
\caption{Effect of capacitance on the 100-meter radius toroid for detection of the passing charge $Q_\mathrm{b}$ using $L=20$~pH.  The nominal value of $R$ is described in the text to correspond to quality factor, QF=10.\label{charging-phase}}
\end{figure}

Now consider a much larger device, with a  a cosmic-ray air shower passing through it.  We take its dimensions to be extremely large, with a 100~m inner radius.   We take the outer radius to be 0.1~m larger and the width to be 0.1~m.  Because the inductance drops as $1/r$ and is 20~pH, we use a larger capacitance, 5~$\mu$F  to meet our requirements (the conditions for $I=I_b$). The impedances of the inductor and capacitor are
0.004~$\Omega$ and 0.001~$\Omega$, respectively.     The impact of the choice of capacitance on its maximum stored charge relative to the ideal value, $Q_{\mathrm{b}}$, is shown in Figure~\ref{charging-phase}.  For this figure, the simulated impulse was a 5~ns rectangular pulse low-pass filtered at 300~MHz to remove sharp edges in the plot.  (The plots look essentially the same with full bandwidth.)   The resistance of the copper is negligible, but there is an effective resistance that sets the quality factor QF of the resonator.  Choosing QF to be an easily achievable value of 10, we select resistances in the range 0.0002 to 0.002~$\Omega$.  The results of this charging phase are barely sensitive to any reasonable values of QF.

We also consider an even slower large detector so that we can use the higher magnetic permeabilities at lower frequencies and even slower electronics, and take $C=5$~mF as a comparison point.
Along the way, we may want to prototype a smaller, $r=1$~m,  version with otherwise the same values for $d$ and $w$.  Choosing $C=0.05$~$\mu$F, the results for the inductor, resistor and capacitor respectively are: $0.4$, $0.1$, $2\times 10^{-4}$~$\Omega$.
These results and other values are shown in Table~\ref{toroidparams}.
\begin{table}
\begin{center}
\begin{tabular}{|l|l|l|l|l|l||l|l|l|}
\hline
application & $r$ & $d$ & $w$ & $L$ & $C$ & beam& $|Z_L|$ & $|Z_C|$ \\
& &  &  & & &pulse&  & \\
 & (m) & (m) & (m) & & & (ns) & ($\Omega$) &($\Omega$)  \\
 \hline\hline
 beams &  0.05 & 0.01 & 0.025 & 1 nH & 4000 pF & 0.01 & 12000 & 0.2 \\
 1-m prototype & 1.0 & 0.1 & 0.1 & 2 nH & 0.05 $\mu$F & 5 & 0.4 & 0.1 \\ 
 Cosmic Rays (fast) & 100 & 0.1 & 0.1& 20 pH & 5 $\mu$F & 5 & 0.004 & 0.001 \\
 Cosmic Rays (slow) & 100 & 0.1 & 0.1& 20 pH & 5 mF & 5 & 0.004 & $10^{-6}$ \\
 \hline
 \end{tabular}
 \caption{Impedances of the toroidal detector for different geometries and magnetic field duration\label{toroidparams} during the charging phase.}
 \end{center}
 \end{table}

%We check the above arguments analytically using a rectangular impulse for the beam current, $I_b(t)$ and plot the voltage on the storage capacitors.   Because no system is infinite bandwidth, we assume the voltage is measured with a [0--300] or [0--3000]~MHz bandwidth by applying a low-pass filter to our results.  Because we are unable to analytically treat the changing resistance and inductance with frequency, we zoom in here on the early times $<$10~ns, and using the short-duration values such as $\mu_r=1$.  Figure~\ref{charging} shows the peak voltage, relative to the ideal value $V_{ideal}=Q_{b}*C$ where $Q_{b}=\int I_{b} dt$.  In all cases, $V_{peak}$ times the capacitance yielded the theoretical charge within 1\%.  
%During the discharging phase, we will observe that much smaller bandwidths are actually needed than considered here.
 
The larger capacitance unfortunately lowers the minimum detectable charge, because $V=Q/C$.  However, it also makes the discharging phase longer  so there is less bandwidth for noise.  Also simple low-speed electronics can then be used to measure the charge.  We examine such tradeoffs for pulse shapes and noise further in the following subsection.

If the passing beam is off-center, the above analysis should still apply.  A typical ICT storage capacitor is really about 40 small parallel capacitors distributed around the toroid with one overall effective capacitance and corresponding stored charge.  Upon the initial charging the capacitors will not charge equally but we expect them to equilibrate on time scales much shorter than seen in the discharging phase. There may be some losses which need to be investigated experimentally, but are not seen in implementation by the commercial ICT.

\subsection{Discharging Phase}

Now that the capacitor is charged we need to know how the voltage changes with time as it discharges.   This allows us to determine how close the peak voltage is to determining the charge $Q$ on the sampling capacitor using the ideal formula $Q=CV$.   The time dependence of the charge on the capacitor is given by the homogeneous version of Equation~\ref{maindiffeqn} rewritten for the charge $Q(t)$:
\begin{equation}
R\frac{dQ}{dt}+L\frac{d^2Q}{dt^2}+\frac{Q}{C} =0.
\end{equation}

\begin{figure}
\begin{center}
\includegraphics[width=5in]{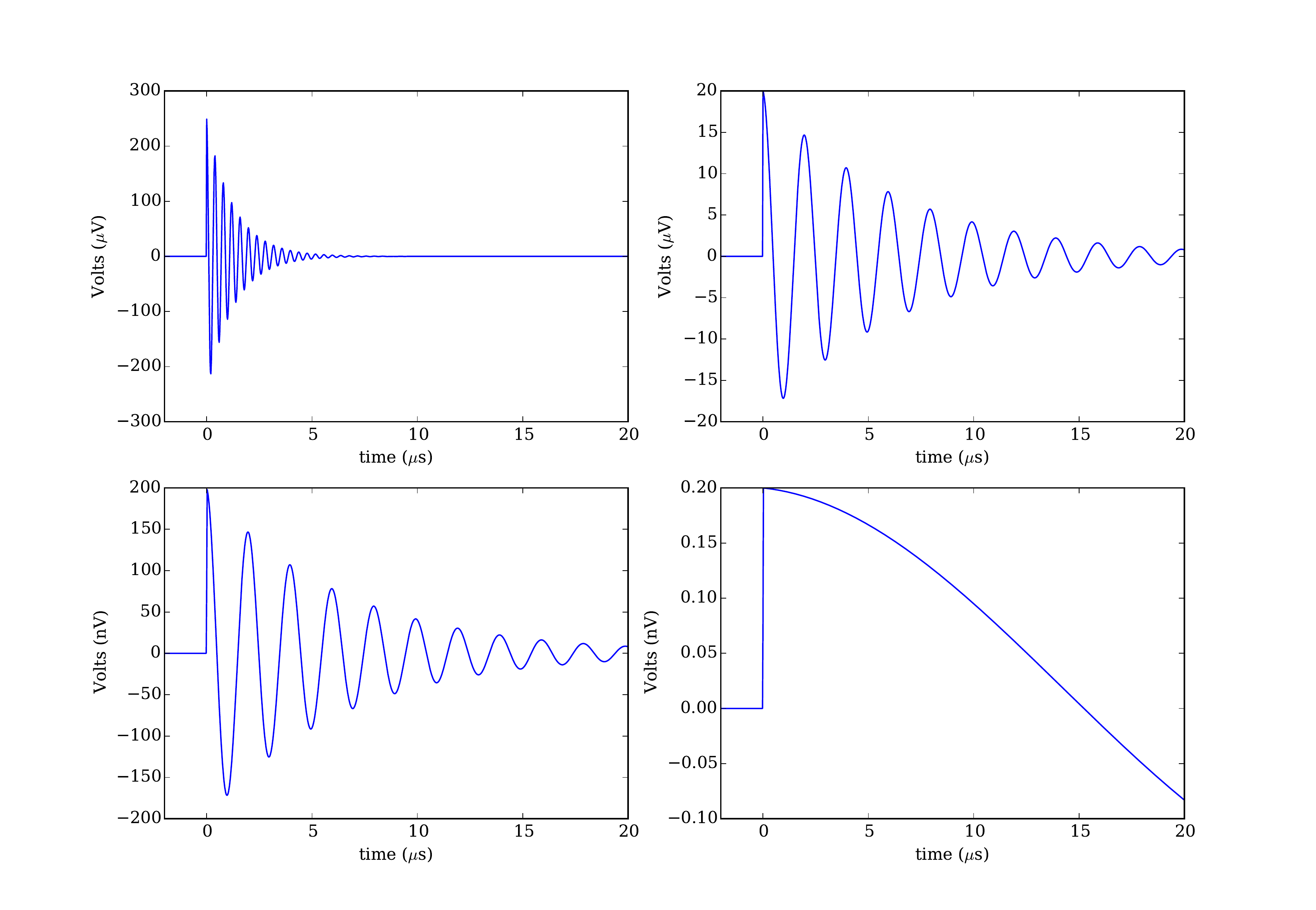}
\end{center}
\caption{Expected output pulses for the four toroidal configurations discussed in the text: a) a small 0.1~m radius device such as used for beam monitoring, b) a one-meter radius prototype, c) a 100 meter radius device with fast discharge time, and d) the same 100 meter radius device with slower discharge. The plots assume a quality factor of the resonator equal to 10.
\label{output-pulses}}
\end{figure}

The discharge is much slower than the charging time because $L$ is larger by a factor of $\mu_r$, the relative permeability.  For an initial charge  $Q_0$ on the storage capacitor, the solution $V(t)=Q(t)/C$ in the underdamped case is
\begin{equation}
V(t) = -\frac{Q_0}{C}[\cos(\omega_0 t)- \frac{R/2L}{\omega_0}\sin(\omega_0 t)]
e^{-\frac{R}{2L}t},
\end{equation}
where $\omega_0\equiv \sqrt{\frac{1}{LC}-\frac{R^2}{4L^2}}$.
For underdamping we require $\zeta \equiv \frac{R}{2}\sqrt{\frac{C}{L}} < 1$ which is the case for all physical scenarios for discharging that we consider in this article.
 We simulate the discharge using an example core material  with low-frequency permeability $\mu_r=1000$.    We note that much larger values may be achievable at lower frequencies and/or with new materials.
 Figure~\ref{output-pulses} shows the expected discharge curves for the various devices considered above.

The exact decay time of the envelope of the discharge curves will be determined by losses, resistive as well as due to the imaginary parts of the dielectric and magnetic materials.  Our estimates of resistive losses are much lower than the latter two, where magnetic losses probably dominate over dielectric.   We chose a quality factor of 10 to match easily achievable values for $RLC$ resonators.  The long ringing offers the opportunity for multiple measurements as well as narrow-band filtering to remove noise.   The ringing frequency is a property of the device and not its stimulus so a matching filter can even be employed.    We expect that one can extract very small signals using a matched filter in hardware or software via cross correlations with a waveform template.

We note in passing that commercial ICTs do not ring like this, because of the presence of a secondary set of windings which rapidly removes energy from the resonator.
We suspect we can improve the sensitivity of the ICT by measuring the full ringing output to determine the charge.  That is we are dissipating power more slowly, about 100 times longer, which gives more opportunity to measure the signal.  Still, the output is short enough not to interfere with successive beam pulses or cosmic rays.

%The larger device required larger capacitance because the inductance was so much smaller.  We want the oscillation time scale represented by $2\pi/\omega_0$ to be longer than time for currents to travel across the device. This slows the output signal and lowers the bandwidth, thereby reducing noise and making implementation of the electronics easier and less expensive.  A larger $C$ will lower the voltage measured for a given charge and we investigate those tradeoffs below.

\subsection{Sensitivity}

The fundamental sensitivity of the device will be set by the thermal noise on the capacitor which causes a variance of its charge and corresponding voltage.  The charge on a capacitor may be treated as one degree of freedom of a simple harmonic oscillator and so the standard deviation of the charge, $\delta Q$, is given by its one degree of freedom  $\frac{1}{2}\frac{(\delta Q)^2}{C}=\frac{1}{2}k_B T$, where $k_B$ is Boltzmann's constant and we assume $T=290$~K.  Hence $\delta Q=\sqrt{k_B T C}$, which is called kTC noise.     

For the case of the beam ICT, with $C=4000$~pF, $\delta Q$ is 4~fC.  Thus thermal noise on the capacitor is not the limiting noise source for the beam ICT which, as far as we can tell is currently limited to about 300~fC by other sources.  For the 100~meter versions we consider here, the fast device has $C=5$~$\mu$F and so $\delta Q=0.141$~pC (884k electrons).   The slower device with $C=5$~mF would have $\delta Q=4.5$~pC (28M electrons).   As mentioned above, because of the long ringing structure of the output pulse, we may have the opportunity to lower this noise estimate even further and will be determined experimentally in future work.

To consider further sources of noise, we need to know the system bandwidth.  We note that the output time scales are determined by $\sqrt{LC}$.  The self-inductances $L$ were given in Table~\ref{toroidparams} for the fast-charging phase and assumed relative permeability of one. On discharge, which is slower, we can easily achieve relative permeabilities of $\mu_r=1000$ with a corresponding increase in $L$.  Hence the timescales for the devices are 63~ns (ICT), 316~ns (fast cosmic ray detector) and 10~$\mu$s (slow cosmic ray detector)

%The noise of the system will include noise of the front-end amplifier which includes both current noise and voltage noise.  We want the input impedance of the amplifier to be larger than the impedance of the capacitor when it is discharging.  For the ICT, the discharging $\omega$ corresponds to $\omega \sim$~1/(50~ns) so the capacitor's impedance is $1/(\omega C) \sim$~13~$\Omega$.  Hence for an ICT even a $50~\Omega$ input impedance to the amplifier just about suffices.   50 ohm noise calc

%For the larger cosmic ray detectors, $1/\omega$ is scaling as $\sqrt{LC}$ so $Z_C=1/(\omega C)$ scales as $1/\sqrt{C}$.
%So, the larger capacitance of the cosmic detectors  requires a larger input impedance.  In each case a 1~k$\Omega$ should suffice.  We consider an example high quality front-end amplifier chosen for this input impedance, Analog Devices AD8099ARDZ with voltage noise 0.95~nV/$\sqrt{\mathrm{Hz}}$ and current noise 2.6~pA$\sqrt{\mathrm{Hz}}$.  For 1k$\Omega$ input impedance and bandwidths for the fast and slow cosmic ray detector of 3~MHz and 100~kHz we have noise values of 5~$\mu$V and 876~nV respectively.  The corresponding $\delta Q$ is 25~pC and 4380~pC, respectively.  These results are summarized in Table~\ref{noiselevels}.  Hence the front electronics serves as the limiting factor, and there should be room for improvement.
\begin{table}
\begin{center}
\begin{tabular}{|l |l |l |l |l |l |l|}
\hline
application & $C$ & L & kCT  & kCT &   sensitivity & inverse \\ 
 & & & noise & noise &   & bandwidth\\
     &    & &  (pC) & ($N_{\mathrm{ch}}$) & (nV) & $\sqrt{LC}$ (ns)\\
\hline\hline
ICT & 4 nF &  1$\mu$H & 0.004 & 25000& 1000 & 63 ns\\
1-m prototype & 0.05~$\mu$F & 2$\mu$H &0.014 & 87,500 & 290 & 316 ns\\
Cosmic Rays (fast) & 5~$\mu$F & 20nH &0.14 & 875,000  & 29 & 316 ns\\
Cosmic Rays (slow) & 5~mF & 20nH &4.5 & 28M & 0.91 & 10 $\mu$s\\
\hline
  \end{tabular}
\caption{Sensitivity for detection of relativistic charges for the four different devices considered.\label{noiselevels}}
\end{center}
\end{table}
For this discussion we are assuming we can chose an amplifier whose internal voltage and current noise are negligible compared to kCT noise.  The numbers presented in Table~\ref{noiselevels} beyond the ICT are challenging to achieve electronically.  However, we have not included the effect of multiple measurements which can be applied throughout the decay time.   
%We have not included the fact that we can achieve higher relative permeabilities, perhaps $10^5$ and so slow down the signal without decreasing the peak value.

%Barkhausen noise (why is this only for high R?)

One can also be concerned with the effect of external fields on the device.  Purely magnetic shielding would be difficult to achieve.  
However, to charge the detector's capacitor, the stray fields would need to be  of extremely short duration, on the scale of nanoseconds to picoseconds, since $\omega$ being large is one of our conditions for $I$=$I_b$.  
So, if care is taken not to be in the near field of any fast magnetic sources, any other fast magnetic fields will arrive as part of electromagnetic waves.    The entire detector could be sandwiched or encased in conductive mesh which removes this energy by dissipating its electric component.  
Furthermore, any continuous-wave (CW) source will produce a substantially equal amount of both signs of magnetic field when averaged over the integration time of the device.    The only way to make a true pulse of magnetic field would be from a relativistic current, which is exactly what the detector is looking for.  Other stray electromagnetic fields could interfere with the measurement of the capacitor's stored charge, for example by being picked up directly by the amplifier.  However, such noise sources are typically of a characteristic frequency which can be notch filtered.   

We are not overly concerned with signals induced by other moving charges such as atmospheric electricity since they are not relativistic.  Slow-moving charges do not make magnetic fields short enough to satisfy the fundamental requirement that the impedance of the inductor be larger than that of the capacitor.

\section{Other geometries}

The detector can also be built in other geometrical configurations. For example, we can replace the toroid by a regular polygon, which is easier to manufacture. Moreover, we can deploy only one side of this polygon, making the detector into a rod of length $l$, width $d\ll \ell$, and height $w$. In this case, if a beam passes by at a perpendicular distance $r$ from this rod, with $r\gtrsim l$, the rod can be approximated as an arc of a toroid of radius $r$. As derived in~\ref{inductances}, the mutual inductance between the beam and the rod will be, approximately,
\begin{equation} \label{eqn:rodM}
M = \frac{\mu_r \mu_0 w}{2\pi} \ln (\frac{r+d}{r}).
\end{equation}
When a current $I$ flows around the shell of the detector, neglecting boundary effects, the magnetic field inside the core will be
\begin{equation}
B = \frac{\mu_r \mu_0 I}{l},
\end{equation}
and the magnetic flux will be
\begin{equation}
\Phi = Bwd = \frac{\mu_r \mu_0 I w d}{l}.
\end{equation}
Thus, the self-inductance of the rod is
\begin{equation} \label{eqn:rodL}
L \equiv \frac{\Phi}{I} = \mu_r \mu_0 \frac{w d}{l}.
\end{equation}
Since $r \gg d$, from Equations~\ref{eqn:rodM}~and~\ref{eqn:rodL} we obtain
\begin{equation}
\frac{M}{L} \approx \frac{l}{2\pi r}.
\end{equation}
Thus, by Equation~\ref{eqn:I I_b}, when $|Z_L| \gg R$ and $|Z_L|\gg |Z_C|$ we have
\begin{equation}
I = \frac{l}{2\pi r} I_b,
\end{equation}
so the signal will be scaled by a factor of $\frac{l}{2\pi r}$. If the beam current spreads out in an area $A$ with current density $J_b(\mathbf{r'})$, the induced current will be
\begin{equation}
I = l \int_A \frac{J_b(\mathbf{r'})}{2\pi |\mathbf{r'}|} d^2\mathbf{r'}.
\end{equation}

Compared with the toroidal model, the rod model has a simpler geometry, so it is easier to build and to analyze. Another advantage of the rod model is that it does not predefine the area of detection. If a beam passing by has a small distance of closet approach ($\sim$100~m), it may be detected even if it has a large zenith angle.  However, a single rod cannot tell the difference between weak nearby currents and strong currents farther away.

\section{Cosmic-ray detection}

Cosmic rays offer both an application and test signal for this detector since a narrow pancake of relativistic charge will traverse the plane of the toroid.   At shower maximum for an ultra-high energy air shower, the typical electron energy is 50-100 MeV, corresponding to a boost factor, $\gamma$, of 100-200.   Thus the criterion of a narrow cone of magnetic field is met.   If the shower is inclined, as is likely, the response of the toroid will not be synchronized at all azimuthal points, but we are only considering designs where this effect is still shorter than the discharge time.   

We postpone for future work a full Monte Carlo simulation of event rates and thresholds which would be required for precise event rate estimates.  We show here instead the plausibility of a reasonably high detection rate.   For example, we have shown that the threshold of the 100 meter ``fast'' cosmic ray detector is 875,000 relativistic electrons (see Table~\ref{noiselevels}).   For a clear signal that is well above background, we estimate the energy of an air shower corresponding to a signal/noise ratio of 5.  In addition we note that the detector only responds to charge excess, $(N_{e^-}-N_{e^+})/(N_{e^-}+N_{e^+})$, which is typically about 20\% in air showers.~\cite{askaryan1}  As a result we use the primary energy threshold corresponding to $(1/0.2)\times 5\times 875\mathrm{k}$=22 million charges near shower maximum.

The number of relativistic charges from a cosmic ray can be estimated from Ref.~\cite{grieder} (figure 6.2a) to be 20M charges for a 100 PeV shower at shower maximum.   We will operate at an altitude for which the median such shower intersects the ground. However, even after considering variations due to shower fluctuations and different inclination angles, enough charges will reach the ground to be detected.

The ideal altitude for placing such a detector depends on the range of zenith angles, $\theta_z$, that are considered since that will affect the slant depth, {\it i.e.}, the material overburden measured  in g/cm$^2$.  We consider the typical useful range of $\theta_z$=0 to 60$^{\circ}$  by surface detectors.  Based again on Ref.~\cite{grieder} (figure 6.2a), 100~PeV showers over this zenith range will intersect with the ground at shower max for an atmospheric depth of 700~g/cm$^{2}$.   According to Ref.~\cite{grieder} (equation 6.18), the relationship of vertical column density, $X$, to height above sea level, $h$, is 
$X(h)=(1030~\mathrm{g/cm^{2}})\cdot \exp(-h/7.3~\mathrm{km})$. For example, at the HAWC~\cite{hawc} 
or ARGO-YBJ~\cite{argo-ybj} detectors location of $\sim$4.2~km altitude, the column density range over $\theta_z$=0-60$^{\circ}$ is 580 to 1160~g/cm$^{2}$.  Over this variation the number of charges and ground level ranges from 25-100\% of shower max, well within our threshold budget.   Such a detector might be useful as a surface veto for large neutrino detectors, such as upgrades to 
IceCube.~\cite{icecube}

We now estimate the event rate of 100~PeV proton-initiated showers intersecting the 100-meter radius detector.  The Moli\`{e}re radius of these showers is smaller than the size of the toroid and we neglect edge effects.    From Ref.~\cite{pdg-cr}, we estimate 
$dN/dE=(7.9\times 10^8) (E/\mathrm{GeV})^{-3.3} \mathrm{GeV^{2.3} m^{-2} s^{-1} sr^{-1}}$.
Over these zenith angles, our 100 meter radius detector projects an average area of $\frac{3}{4}\pi 100^2$~m$^{2}$. So in one year in this zenith angle range (which corresponds to $\pi$ steradians) we expect 300 events, or about one event per day.     Using instead the integral plot from Ref.~\cite{grieder} (figure 11.19) we find 2300 events, or even more per day but is roughly consistent. A full simulation, which will include detection of fluctuations of lower energy showers will probably yield an even higher value. 

\section{Discussion}

In particle astrophysics, one is often faced with the difficult problem of instrumenting large areas.  In this paper we have put forward and analyzed a technique which takes advantage of Amp\`{e}re's law to provide sensitivity over a two-dimensional surface while only instrumenting a one-dimensional contour.    In addition to cost savings by this reduction in dimension, we note that the sensitivity over the enclosed area is 100\%, unlike sampling techniques.   For example, the Pierre Auger Observatory~\cite{pao} covers $3\times 10^9$~m$^2$ but only with extreme sampling.  This distinction would be important for detecting a flux of exotic particles which the sampling technique would miss.  For complete area coverage, one is left to much smaller detectors such as Super-Kamiokande~\cite{superK}, which instruments an area of $1.6\times 10^3$~m$^2$.  Even just one of the 100~m radius detectors considered in this paper would give complete coverage of a larger area, namely $3.1\times 10^4$~m$^2$.     

We have discussed the plausibility of detecting ultra-high energy cosmic rays with this device and achieved promising numbers for detection.   Once the principle is proven experimentally, this device may prove an efficient surface veto for large cosmic neutrino detectors.

The need to instrument only a contour may have implications for space-based detection as well, where weight is a primary concern.  A large area detector could be deployed by a satellite or probe by unfurling the detector we have described.   Such a detector would not only detect fast currents from plasma or other sources, but also be the fastest magnetic field detector yet deployed.

 We have taken a new step in the design used for integrating current transformers (ICTs) deployed at accelerators.  By reading the stored charge on its capacitor directly through its voltage rather than requiring its discharge, is a significant simplification which we suspect from our preliminary laboratory measurements will yield better sensitivity.  Furthermore, this allows the ICT signal to persist for microseconds as shown in Figure~\ref{output-pulses}a . This hundred-fold increase in duration and predictable waveform offers the possibility of a much more sensitive measurement by use of a narrow-band and/or matched filter.  This will be subject of future work.

\section{Acknowledgments}

We thank James Rosenzweig for encouraging discussions and for the loan of an ICT for reverse engineering.    We thank Keith Bechtol for noting the application of this detector technique as a cosmic-ray veto for experiments such as IceCube upgrades. This material is based upon work supported by the Department
of Energy under Award Number DE-SC0009937.

\appendix
\section{Mutual and self inductance calculation}
\label{inductances}

Consider a toroid of inner radius $r$, outer radius $r+d$, and height $w$, as in Figure~\ref{toroid}. Suppose a beam, $I_b$, passes through the center of the toroid. The magnetic field at a distance $r'$ from the beam is
\begin{equation}
B_b(r')=\frac{\mu_r\mu_0 I_b}{2\pi r'} \label{eqn:B_b(r')},
\end{equation}
where $\mu_r$ is the relative permeability of the magnetic material. Therefore, the flux, $\Phi_b$, through a cross-section of the toroid (which is a rectangle with side length $w$ and $d$) is:
\begin{eqnarray}
\Phi_b &= &\int_r^{r+d} \frac{\mu_r\mu_0 I_b}{2\pi r'} w dr'\\
\vspace*{0.5in}\nonumber\\
 & = & \frac{\mu_r\mu_0 w I_b}{2\pi} \ln (\frac{r+d}{r})
\end{eqnarray}
The mutual inductance between the space that is carrying the beam (as if a wire) and the toroid is given by $M \equiv \Phi_b/I_b$. Hence,
\begin{equation}
M = \frac{\mu_r\mu_0 w}{2\pi} \ln (\frac{r+d}{r}) \label{mutual inductance}
\end{equation}

To calculate the self-inductance of the toroidal shell, we assume that there is a current, $I$, flowing around the cross-section of the toroid, and it is azimuthally symmetric. By symmetry, the magnetic field inside the toroid points in the azimuthal direction, and its magnitude depends only on the distance from the center, $r'$ ($r<r'<r+d$). Taking a loop of radius $r'$, then by Ampere's law we have
\begin{equation}
\mu_r\mu_0 I = 2\pi r' B(r')
\end{equation}
Therefore
\begin{equation}
B(r') = \frac{\mu_r\mu_0 I}{2\pi r'} \label{eqn:B(r')}
\end{equation}
Note that this is the same as Equation~\ref{eqn:B_b(r')}. Therefore, we can obtain the magnetic flux through a cross-section:
\begin{equation}
\Phi = \frac{\mu_r\mu_0 w I}{2\pi} \ln (\frac{r+d}{r})
\end{equation}
Thus, the self-inductance of the toroidal shell, defined by $L\equiv \Phi/I$, is
\begin{equation}
L = \frac{\mu_r\mu_0 w}{2\pi} \ln (\frac{r+d}{r}) \label{self-inductance}
\end{equation}

Comparing Equations~\ref{mutual inductance}~and~\ref{self-inductance}, we can see that $M$=$L$.

The derivations above assume azimuthal symmetry (i.e. the beam passes through the center of the toroid, and the current on the shell is evenly distributed around the toroid). In the more general case, the results are not exact. However, because we use magnetic materials with high permeability (and thus low magnetic reluctance), magnetic flux is mostly preserved inside the toroid. Therefore, under approximation we can still assume that the magnetic field inside the toroid depends only on the distance $r'$ from the center. Since the current enclosed in the loop of radius $r'$ is still $I_b$ for the beam current and $I$ for the current in the copper shell, Equations~\ref{eqn:B_b(r')}~and~\ref{eqn:B(r')} still hold, so Equations~\ref{mutual inductance}~and~\ref{self-inductance} are still valid.

\section*{References}

\bibliography{mag-det}

\end{document}